\documentclass[conference]{IEEEtran}
\ifCLASSINFOpdf
\else
\fi
\usepackage{flushend}
\usepackage{graphicx}

\usepackage{cite}
\usepackage[linesnumbered,lined,ruled,commentsnumbered]{algorithm2e}
\usepackage{amsmath}

\begin{document}
\bstctlcite{IEEEexample:BSTcontrol}
\title{ A Cache Reconfiguration Approach for Saving Leakage and Refresh Energy in Embedded DRAM Caches}
%
%
%

%
%

 \author{\IEEEauthorblockN{Sparsh Mittal }
 \IEEEauthorblockA{Department of Electrical and Computer Engineering \\
 Iowa State University, Ames, Iowa 50011, USA\\
  Email: sparsh0mittal@gmail.com}
 }


\maketitle

\begin{abstract}
     
In recent years, the size and leakage energy consumption of large last level caches (LLCs) has increased. To address this, embedded DRAM (eDRAM) caches have been considered which have lower leakage energy consumption; however eDRAM caches consume a significant amount of energy in the form of refresh energy. In this paper, we present a technique for saving both leakage and refresh energy in eDRAM caches. We use dynamic cache reconfiguration approach to intelligently turn-off part of the cache to save leakage energy and refresh only valid data of the active (i.e. not turned-off) cache to save refresh energy. We evaluate our technique using an x86-64 simulator and SPEC2006 benchmarks and compare it with a recently proposed technique for saving refresh energy, named Refrint. The experiments have shown that our technique provides better performance and energy efficiency than Refrint. Using our technique, for a 2MB LLC and 40$\mu$s eDRAM refresh period, the average saving in energy over eDRAM baseline (which periodically refreshes all cache lines) is 22.8\%.

\end{abstract}




%
\IEEEpeerreviewmaketitle


%


\section{Introduction}
To fulfill the performance requirements of state-of-the-art resource-intensive 
applications (e.g. \cite{4554309}), while also meeting the constraint posed by 
chip power budget, achieving high energy efficiency has become vital in modern 
processor design to continue to scale performance 
\cite{Mit_DRAMsurvey,pande2013demonstration}. To bridge the gap between the 
speed of processor and memory, modern processors use large last-level caches 
(LLCs) which occupy a large fraction of chip area. Further, due to large leakage 
energy consumption accompanying CMOS technology scaling 
\cite{rodriguez2006energy}, the power consumption of SRAM-based LLCs has become 
a significant fraction of processor power consumption 
\cite{monchiero2006design}. In Niagara-2 processor, L2 cache consumes nearly 
25\% of the total power consumption \cite{li2009mcpat}. Thus, improving the 
energy efficiency of LLCs has become an important issue in modern processor 
design. 

Recent advances in chip fabrication have enabled use of embedded DRAM (eDRAM) to build on-chip caches, for example, the last level caches in IBM's Power 7 processor \cite{kalla2010power7} and Blue Gene/L supercomputer chip \cite{iyer2005embedded} are designed using eDRAM. Compared to conventional SRAM-based caches, eDRAM caches offer higher density and lower-leakage. A crucial limitation of eDRAM cells, however, is that they lose charge over time and hence require refresh operations. The retention period of an eDRAM cell is defined as the duration of time for which the cell can retain its state; thus to avoid failures, the cell needs to be refreshed before its refresh period. Since eDRAM uses fast logic transistors with a higher leakage current than the  conventional DRAM, the refresh period for eDRAM is nearly a thousand times shorter than that of conventional DRAM. For example, Barth, et al. \cite{barth2008500} report the refresh period of eDRAM to be 40$\mu$s, in comparison to the 64ms refresh period of a commodity DRAM \cite{wilkerson2010reducing}. Further, with ongoing technology scaling, increasing leakage and smaller storage capacitance further reduce the retention period which require more frequent refresh, thus leading to larger power consumption \cite{changtechnology}. It has been shown that for eDRAM LLCs, refresh energy accounts for nearly 70\% of the total energy consumption, while the leakage energy accounts for most of the remaining fraction \cite{agrawalrefrint}. Thus, reducing the power consumption of eDRAM is extremely important for fully leveraging their features and enabling wide-spread use.
     
In this paper, we present a technique for saving energy in eDRAM LLCs. Our technique saves both leakage and refresh energy in eDRAM caches. It uses dynamic cache reconfiguration approach to intelligently turn-off part of the cache to save leakage energy and refreshes only \textit{valid} data of the active (i.e. not turned-off) cache to save refresh energy. Since the active portion of the cache is generally reduced, the refresh requirements are also reduced, which leads to reduction in the refresh energy of the cache. Our technique uses low-cost dynamic profiling to collect information about the running program  and uses an algorithm to periodically compute the energy consumption of different possible cache configurations. Using this, the cache is reconfigured to the most energy-efficient configuration. Our technique does not require offline profiling or manual tuning of its parameters. The energy saving algorithm runs in software and uses lightweight hardware support. By virtue of this, the algorithm can easily take into account, components of the processor other than cache, such as main memory etc.

We evaluate our technique using Sniper x86-64 simulator and SPEC2006 benchmarks and compare it with a recently proposed technique for saving refresh energy, named Refrint polyphase-valid (RPV)  \cite{agrawalrefrint}. The experiments have shown that our technique provides better performance and energy efficiency than RPV. For 40$\mu$s refresh period and a baseline 2MB eDRAM LLC (which periodically refreshes all cache lines), the energy saving achieved using our technique and RPV are 22.8 and 13.5, respectively. Moreover, the experiments performed with 30$\mu$s refresh period show that with decreasing refresh period, the advantage of our technique increases even further. This shows the effectiveness of our technique. We also compare our technique against SRAM LLC (which does not use any energy saving technique) to show the energy saving achieved by  our technique compared to an SRAM LLC. Our technique is expected to be especially useful in resource-constrained environments \cite{kumar2011distributed,mittal2010content}.

The rest of the paper is organized as follows. Section \ref{sec:background} discusses related work in use of eDRAM caches and power management in eDRAM caches. Section \ref{sec:methods} discusses the components of our technique. Section \ref{sec:hardwareimplementation} discusses the hardware implementation and Section \ref{sec:experimentmethodology} discusses the simulation methodology. Section \ref{sec:results} presents the experimental results and parameter sensitivity analysis. Finally, Section \ref{sec:conclusion} presents the conclusion and future work.

%




\section{Background and Related Work}\label{sec:background}
Modern computing systems such as laptops, desktops, servers and other portable 
systems are currently being employed in a wide range of applications 
\cite{pande2010algorithms,MitMit2011_QA,pande2007network}. However, as the 
carbon footprint of IT increases, the emphasis on energy efficiency in the 
design of computing systems is expected to increase further. To address this 
challenge, researchers have proposed device-level and architecture-level 
innovations and techniques.  In recent years, use of low-leakage devices such as 
eDRAM has become a promising approach to manage power consumption of LLCs.

Researchers have proposed several techniques for reducing energy consumption of eDRAM caches and DRAM-based main memory systems. Chang et al. \cite{changtechnology} propose a dynamic dead-line prediction scheme to reduce eDRAM refresh power. Their scheme avoids refreshing a cache line if the line holds data which are unlikely to be reused. Wilkerson et al. \cite{wilkerson2010reducing} propose using error-correcting codes to dynamically detect and correct bits that fail, which allows using longer refresh periods for saving energy. For DRAM systems, Ghosh and Lee \cite{ghosh2007smart} proposed a Smart-Refresh technique which avoids refreshing the DRAM rows which are recently read or written. However, Smart-Refresh approach does not save energy when the intensity of access is low.  Reohr \cite{reohr2006memories} discusses several approaches for refreshing eDRAM caches, for example, no-refresh, periodic refresh and line-level refresh based on time stamps. 
Wu et al. \cite{wu2009hybrid} propose hybrid cache architectures, where different levels of cache in the cache hierarchy or different regions of the same cache can be made of disparate memory technologies, such as eDRAM, phase change memory etc.

Several authors have proposed cache reconfiguration based leakage energy saving 
techniques in the context of SRAM caches 
\cite{HanHri02_TVLSI,YanPow01_IcacheResize,mittal2013PhDThesis}. To the best 
of our knowledge, this is first work which uses cache reconfiguration for 
targeting both leakage and refresh energy in eDRAM caches.

\section{Methodology}\label{sec:methods}

In this paper, we assume that the LLC is an L2 cache, and based on the explanation here, our technique can be easily shown to work for the case when the LLC is an L3 cache. We use the term `interval' to denote the time after which algorithm is executed and `refresh-event' to denote the event when refresh signal is sent to refresh all or selected blocks of the cache. Note that since interval length is larger than the period of refresh-events, several refresh-events may happen within an interval.

\subsection{Main Idea}

Our technique works on the observation that there exists large intra-application and inter-application variation in cache demands of different applications. Based on this, our technique  dynamically changes the active cache size of an application, such that the performance is minimally affected while a large energy saving is obtained.  As for refresh, our technique periodically refreshes only valid lines in the cache. Our technique reduces the number of refreshes due to three reasons. First, it avoids refreshing invalid blocks. Second it reduces the number of active and valid blocks inside the cache. Third, the reduction in execution time due to smaller number of refreshes also reduces the number of periodic refresh-events required.

\subsection{Cache Coloring}
For selectively allocating cache space to the application, we use cache coloring 
scheme \cite{KesHil92_PageColoring,mittal2013cache}, which we briefly summarize 
below. In this scheme, the cache is logically divided into multiple (say $M$) 
non-overlapping bins, called cache colors. Further, the physical pages are also 
divided into $M$ memory regions based on the least significant bits (LSBs) of 
their physical page number. 
 
The maximum number of colors, $M$ is given by 
\begin{equation} \label{eq:Nvalue}
M = \dfrac{S_{L2}}{ P \times W} \end{equation}
Here $S_{L2}$ shows the L2 size, $P$ shows the system page size (= 4 KB in this paper) and $W$ shows the L2 cache associativity.

Cache coloring scheme maps every memory region (and hence, all physical pages in 
that region) to a unique color in the cache \cite{mittal2013cache}. To record 
the mapping between memory region and cache color, a small mapping table is 
used. By controlling this mapping, such that all the memory regions are 
allocated to only a few cache colors, the cache quota allocated to the program 
can be dynamically controlled \cite{mittal2013cache}. Further, the remaining 
colors are effectively not used and hence, they can be turned-off for saving 
cache energy. To achieve this, at any point of execution of the program, if $m$ 
($<M$) colors are allocated to the program, the mapping table records the 
mapping of $M$ memory regions to $m$ cache colors \cite{mittal2013cache}. Thus, 
cache reconfiguration is achieved at the granularity of a single cache color.

To see the typical size of number of colors, we take the example of a 2MB, 8-way cache with 64B block size. Then from Equation~\ref{eq:Nvalue}, we get $M$ = 64 colors. Thus, the mapping table has 64 entries, each of which is 6-bits (= $\log_2 64$) wide.

\subsection{Profiling Cache}
To estimate the miss-rate for different cache sizes, we use multiple profiling 
units which store auxiliary tags. The profiling units work on the idea of 
set-sampling \cite{puzaksampling,mittal2013PhDThesis}. We use five profiling 
units, each of which profiles size 1X, X/2, X/4, X/8 and X/16 of the L2 cache 
size.  For example, if the L2 cache size is 2MB, the profiling unit marked as 1X 
emulates  an L2 cache of size 2MB and thus provides an estimate of number of 
misses of 2MB cache. Similarly, the profiling unit marked as X/2 provides an 
estimate of number of misses of an L2 cache of size 1MB and so on. The profiling 
units use a large sampling ratio (e.g. 1/64) and do not store data and hence, 
their overhead is small. For example, for a sampling ratio of 1/64 and tag size 
of 30 bits, the overhead of all profiling units is only 0.16\% of the L2 cache 
size.

\subsection{Use of CPI Stack}
For estimating the execution time under different cache sizes, we use CPI stack technique \cite{EyeEec06_CPIOutOrder,CarHei2011_Sniper}. Out of various components of CPI stack, we use memory stall cycle component. We assume that, in an interval, the number of memory stall cycles vary linearly with the number of load misses. Also, we use extra counters in profiling units to also record load misses for different cache sizes. Further, by using the above assumption, we can get memory stall cycles for different load miss values and hence, for different cache sizes. Using this, the execution time under different cache sizes can be estimated.

\subsection{Estimating the Refresh Energy}
 To estimate the refresh energy under different configurations, prediction of the number of cache lines in a refresh period under those configurations is required.  For this purpose, we assume that during an interval the working set of the application remains nearly the same. Our technique maintains a counter $nValid$ to count the number of valid lines present in the cache at any time. The $nValid$ counter is incremented on insertion of a block in the cache and decremented on eviction of a valid block, and thus, the counter is obtained without scanning the array. If $Lines(C_s)$ shows the number of total cache lines at a cache size $C_s$, then the number of refresh operations ($R$) at $C_s$ can be estimated as $R = min(nValid, Lines(C_s))$. By estimating the time consumed in an interval for any cache size, the number of refresh periods can be estimated; and by multiplying it with $R$, the total number of lines refreshed in an interval can be estimated.

%


\section{Energy Saving Algorithm}\label{sec:esa}

In this section, we discuss our energy saving algorithm, which can be a kernel module. The algorithm  is executed after every few million (e.g. 10 million) instructions. In each execution, the algorithm executes the following steps.
\begin{enumerate}
\item First, a set of candidate configurations ($Space_{C}$) is selected using the following criterion.
\begin{itemize}
\item To avoid cache thrashing, at least $C_{Min}$ colors are always allocated to the application. In our experiments, $C_{Min}$ is set to $N/16$. 
\item The granularity of cache allocation is set to be two colors, since this enables us to cover a large space of configurations, while still keeping the algorithm overhead small. 
\item To keep the reconfiguration overhead small, in each interval, a maximum of  $\Delta$ colors can be turned-on or turned-off. In our experiments, $\Delta$ has been set to 16 colors.
\end{itemize} 
 
Thus, the configuration space $Space_{C}$ consists of the configurations which satisfy the following condition:

$Space_{C} = \{ C_i | C_{Min} \le C_i \le M,  C_i\pmod2=0, |C_i-C_{\star}| \le \Delta \} $ 
\item For each configuration in $Space_{C}$, the algorithm estimates the execution time.  In addition, the algorithm also estimates the execution time of a full-size cache (i.e. one with $M$ colors). Let $T_i$ be the execution time estimate of a configuration $C_i$ and $T_0$ be that of the full size configuration. Then, the percentage extra time ($\delta _i$), that $C_i$ is taking over full-size configuration can be estimated as 
  
  \begin{equation}
  \delta _i = \dfrac{T_i-T_0}{T_0}\times 100
  \end{equation}
  
The algorithm computes $\delta _i$ values for all configurations in  $Space_{C}$ and rejects the configurations for which $\delta _i > \beta$. This is to ensure that the algorithm does not choose those configurations which cause large performance loss.  In our experiments, $\beta$ has been set to 3\%.
  
\item For the remaining configurations in $Space_C$, the algorithm computes the memory subsystem energy (as shown in Equation \ref{eq:totalenergy}) and chooses a configuration with the minimum energy consumption for the next interval.  

\end{enumerate}
 
 Note that in each interval, the algorithm examines only a maximum of $\Delta +1$ configurations and hence, its overhead is small.

 \section{Implementation} \label{sec:hardwareimplementation}
We assume that power-gating of eDRAMs for reducing their leakage energy is achieved by a suitable circuit-level technique, as proposed by several authors \cite{chang200765nm,chun2010logic}.

Reconfigurations are handled as follows. When the number of colors of an application are reduced, the data in those colors are flushed and the regions mapped to those colors are mapped to some other color of the application. When an extra color is allocated to an application, some memory regions of the application which are currently mapped to other colors are mapped to this color and thus, the application starts using this color. 

With our technique, block switching takes place only at phase boundaries and not at critical access path of cache. Our technique requires counters for storing the number of misses to different regions in the profiling cache and energy consumption of different configurations etc. However, since the overhead of counters is much smaller than that of memory subsystem (LLC+DRAM), we ignore the overhead of counters.      

The size of mapping table is extremely small and hence, its access latency and energy consumption are extremely small.  The energy saving algorithm runs after  a large interval and hence, its overhead is easily amortized over the phase length.

  
\section{Experimental Methodology}\label{sec:experimentmethodology}

\subsection{Simulation Platform and Workload}\label{sec:Simulation}
We use Sniper, a state-of-the-art x86-64 microarchitecture simulator, which is based on Pin \cite{CarHei2011_Sniper}.  We use interval core model with 128 entry ROB (reorder buffer), 2.2 GHz frequency and a dispatch width of 4 micro-operations.  All caches use a block size of 64B. Both L1D and L1I are 32KB, 4-way, LRU caches and have a latency of 2 cycles. The L2 cache is a 2MB, 8-way, LRU cache with 12 cycle latency.  The latency of main memory is 154 cycles and memory queue contention is also modeled. Interval length is 10M instructions.

We use all SPEC2006 benchmarks with \textit{ref} inputs. For maintaining clarity in the figures, we use three-letter acronyms of the benchmarks, as shown in Table \ref{tab:shortform}. The benchmarks are fast-forwarded for 10B instructions. Then, each benchmark is simulated for 400M instructions.  

\begin{table}[htbp]
  \centering
  \caption{Name of SPEC2006 Benchmarks and Their Acronyms}
    \begin{tabular}{|l|l||l|l|}
   \hline
    Name  & Acronym & Name  & Acronym \\
   \hline
    astar & ast   & libquantum & lib \\\hline
    bwaves & bwa   & mcf   & mcf \\\hline
    bzip2 & bzi   & milc  & mil \\\hline
    cactusADM & cac   & namd  & nam \\\hline
    calculix & cal   & omnetpp & omn \\\hline
    dealII & dea   & perlbench & per \\\hline
    gamess & gam   & povray & pov \\\hline
    gcc   & gcc   & sjeng & sje \\\hline
    gemsFDTD & gem   & soplex & sop \\\hline
    gobmk & gob   & sphinx & sph \\\hline
    gromacs & gro   & tonto & ton \\\hline
    h264ref & h26   & wrf   & wrf \\\hline
    hmmer & hmm   & xalancbmk & xal \\\hline
    lbm   & lbm   & zeusmp & zeu \\\hline
    leslie3D & les   &       &  \\\hline
  
    \end{tabular}%
  \label{tab:shortform}%
\end{table}%

\subsection{Comparison With Other Techniques}

We take eDRAM L2 cache with periodic refresh-all (i.e. both dirty and clean blocks are refreshed) policy as the baseline. For comparison, we also implement Refrint polyphase-valid (RPV) policy \cite{agrawalrefrint}, which works on the intuition that on a read or a write, the line is automatically refreshed and hence, it need not be refreshed for the duration of one retention period. RPV divides the retention period into number of phases. Each cache line maintains the information about the phase in which it was last updated. Afterwards, to reduce the number of refresh operations, RPV refreshes the line at the beginning of this phase, instead of refreshing at beginning of refresh period itself.  We use RPV with four phases, since this is shown to provide significant energy savings \cite{agrawalrefrint}.

Agrawal et al. \cite{agrawalrefrint} also propose Refrint polyphase-dirty (RPD) policy which eagerly invalidates valid blocks to avoid refreshing them and refreshes only dirty blocks. For applications which have only small fraction of dirty data, RPD policy would very aggressively invalidate almost the whole cache; and for small retention periods (e.g. 40$\mu$s translates to only 40,000 cycles  for a 1GHz processor), RPD will greatly increase the access to main memory. Since future technology generations are expected to have even smaller retention periods, RPD will incur large performance loss and hence, we do not evaluate this. Further, RPV policy has been shown to perform better than another policy proposed by Agrawal et al., namely the periodic-valid refresh policy \cite{agrawalrefrint} and hence, we do not evaluate periodic-valid refresh policy.

\begin{figure*}[htp]
\centering
\includegraphics [scale=0.45] {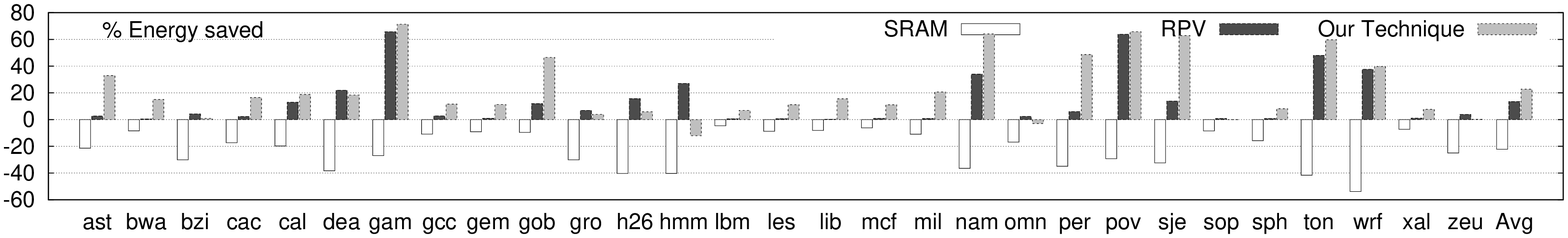}
\includegraphics [scale=0.45] {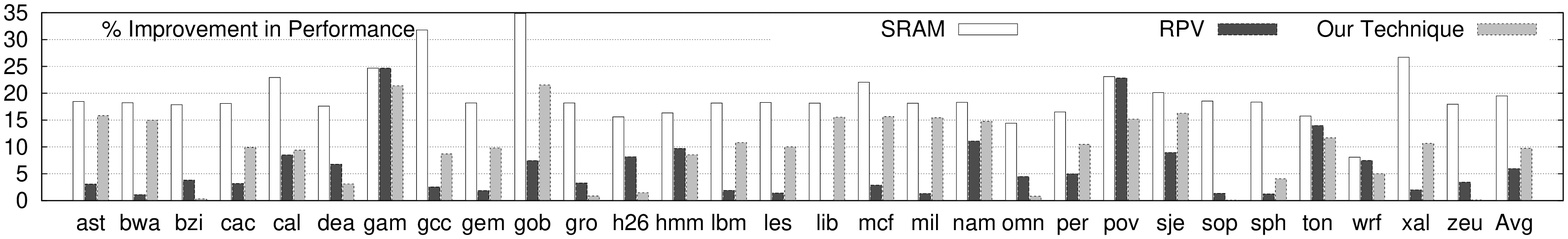}
\includegraphics [scale=0.45] {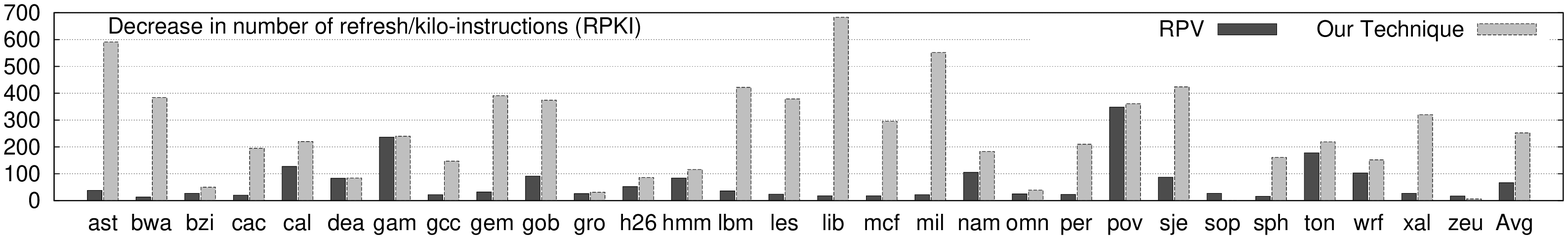}
\includegraphics [scale=0.45] {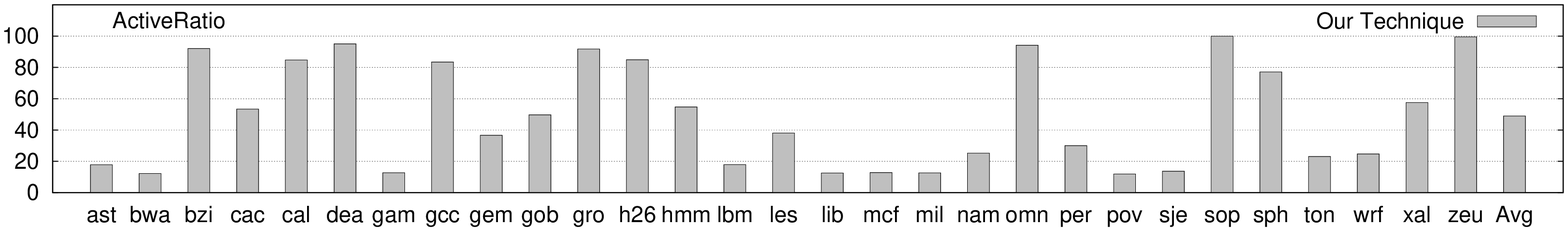}
\includegraphics [scale=0.45] {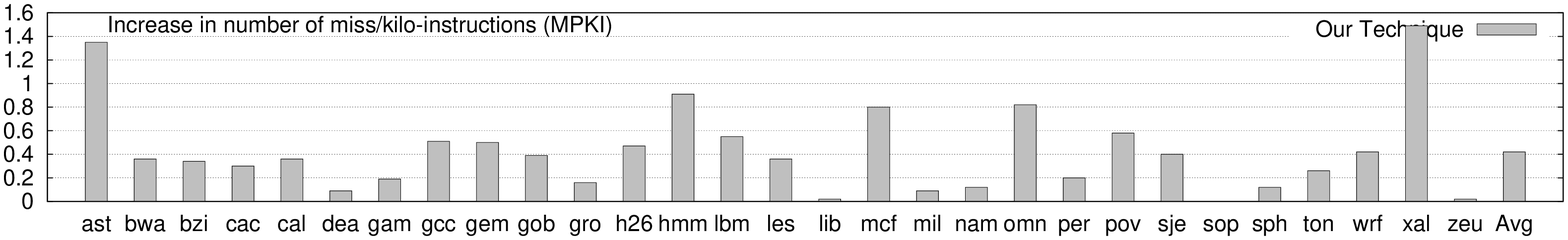}
\caption{Results for different schemes at 40 $\mu$s refresh period. Note that \% improvement in performance refers to \% reduction in simulation time, and hence, a higher value is better. Similarly, for decrease in RPKI and \% energy saved, a higher value is better.}
\label{fig:results40}
\end{figure*}

 \subsection{Energy Model}\label{sec:energymodel}
We account for the energy consumption of L2 cache ($E_{L2}$), main memory ($E_{DRAM}$) and energy cost of algorithm ($E_{Algo}$), since the techniques evaluated here affect the other components only minimally. We use the following notations. $E^{dyn}_{xyz}$ and $P^{leak}_{xyz}$ show the dynamic energy per access and leakage energy per second, respectively, in a component xyz (e.g. L2 or DRAM). For our technique, $B$ shows the number of blocks which are turned on or off; $E_{\chi}$ shows the energy consumed in a single such block transition and $E_{tran}$ shows the total energy consumed in block transitions. $F_A$,  $H_{L2}$ and $M_{L2}$ show the active fraction of cache, number of L2 hits and L2 misses in an interval,  respectively.  $N_R$ shows the number of blocks which are refreshed within all refresh-events in an interval. $T$ denotes the time length of an interval in seconds. $A_{DRAM}$ shows the number of DRAM accesses.  

The L2 leakage energy is assumed to scale with the active fraction of cache 
\cite{mittal2013PhDThesis}. For computing L2 dynamic energy, an L2 miss is 
assumed to consume twice the dynamic energy as that of an L2 hit 
\cite{mittal2013PhDThesis}.  Thus, we have 

\begin{align}
 \label{eq:totalenergy}E&= E_{L2}+E_{DRAM}+E_{Algo} \\
 E_{L2} &= LE_{L2} + DE_{L2} +RE_{L2}  \\   
 LE_{L2} &= P^{leak}_{L2}\times F_{A} \times T \\
DE_{L2} &= E^{dyn}_{L2}\times(2 M_{L2}+H_{L2}) \\
RE_{L2} &= N_R \times E^{dyn}_{L2} \\
E_{DRAM} &= P^{leak}_{DRAM}\times T + E^{dyn}_{DRAM}\times A_{DRAM} \\
E_{Algo} &= E_{\chi}\times B + E_{prof}\\
E_{prof} &= P^{leak}_{prof} \times T + E^{dyn}_{prof} \times A_{prof}
\end{align}

We ignore the energy overhead of RPV algorithm, thus, for experiments with 
baseline eDRAM cache, SRAM cache and RPV, we have $E_{Algo}$ = 0 and $F_A$ = 1. 
We use CACTI \cite{cacti_53} to obtain the values of $E^{Dyn}_{L2}$ and 
$P^{Leak}_{L2}$  at 45nm for SRAM cache, assuming a bank size of 1MB. For 2MB 
cache, we get, $E^{Dyn}_{L2}$ = 0.648 nJ/access and  $P^{Leak}_{L2}$ = 1.296 
Watt. 


Following \cite{agrawalrefrint}, we assume that L2 cache access times and energy 
values are same in both SRAM and eDRAM caches. Further, the leakage energy 
consumption of eDRAM is 1/8th of the SRAM cache 
\cite{iyer2005embedded,agrawalrefrint}. For eDRAM, the time and energy consumed 
in refreshing a line is equal to the time and energy to access the line, 
respectively \cite{agrawalrefrint}. For eDRAM L2, the bank size of L2 is 1MB. We 
assume that each bank of L2 cache  has dedicated logic to process refresh 
requests and using pipelining, a line can be refreshed in a single cycle 
\cite{agrawalrefrint}. 

$ E^{dyn}_{DRAM}$  and $P^{leak}_{DRAM}$ are taken as 70 nJ and 0.18 Watt, 
respectively \cite{mittal2013PhDThesis} and $E_{\chi}$ is taken 
as 2 pJ \cite{mittal2013PhDThesis}. For computing energy values of profiling 
cache, we use CACTI and take the energy consumed in data array only 
\cite{mittal2013PhDThesis}. For a profiling cache, corresponding to 2MB L2, we 
get $E^{dyn}_{prof}$ = 0.0031 nJ/access and  $ P^{leak}_{prof}$ = 0.0050 Watt. 
Clearly, the energy consumption of profiling cache is negligible compared to 
that of L2 cache.

\subsection{Evaluation Metrics}
We take the baseline as an eDRAM cache which periodically refreshes all the 
blocks at the given refresh period. For SRAM L2, RPV and our technique, we show 
the results on following metrics:
\begin{enumerate}
\item Percentage energy saving
\item  Percentage  reduction in execution time
\item Absolute reduction in number of lines refreshed per kilo instructions 
(RPKI). This result is shown only for RPV and our technique.  
\end{enumerate}
For our technique, we also show the results on the following metrics:
\begin{enumerate}
\item ActiveRatio (the fraction of active lines averaged over entire execution 
\cite{mittal2013PhDThesis}) 
\item Absolute increase in MPKI due to use of our technique
\end{enumerate}

 ActiveRatio enables us to evaluate the aggressiveness of cache turn-off of our technique and increase in MPKI helps in evaluating the increase in main memory traffic. Since RPV does not turn-off the cache or cause early invalidation, its ActiveRatio is always 100\% and the increase in MPKI is always zero. Similar is also true for SRAM L2.

\begin{figure*}[htp]
\centering
\includegraphics [scale=0.45] {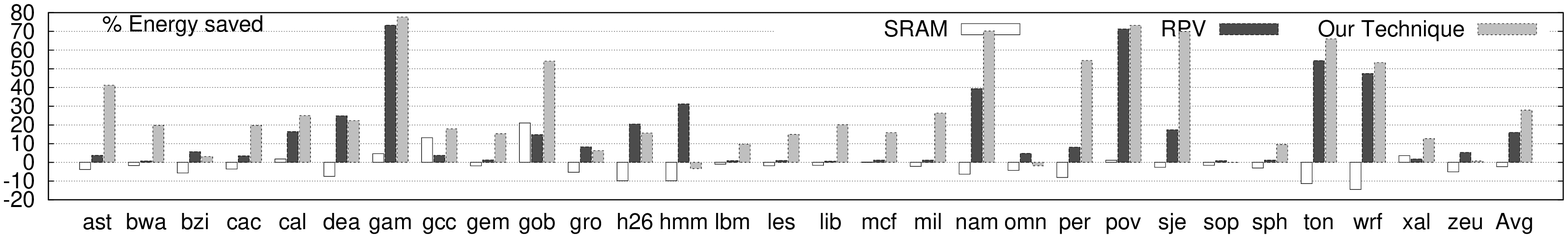}
\includegraphics [scale=0.45] {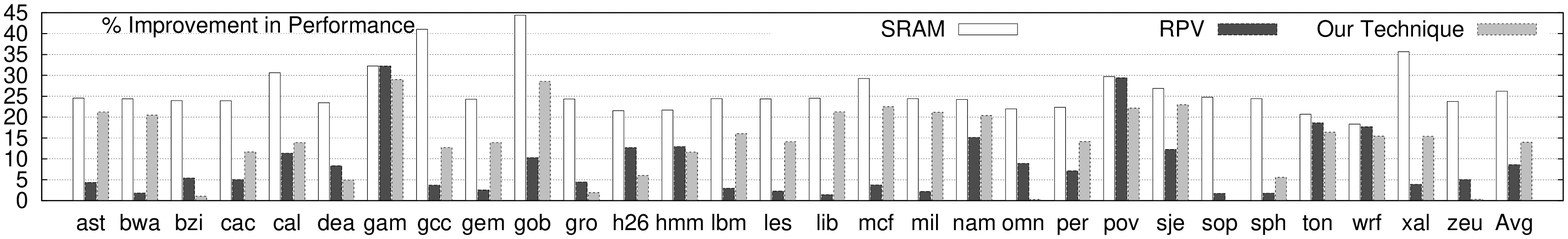}
\includegraphics [scale=0.45] {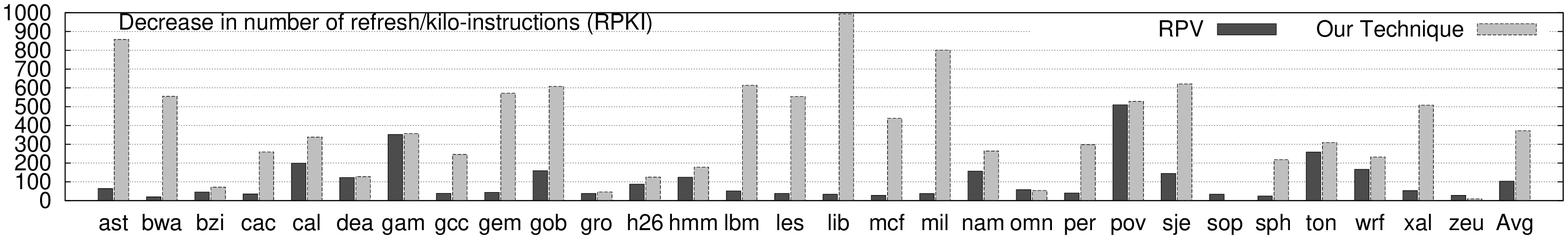}
\includegraphics [scale=0.45] {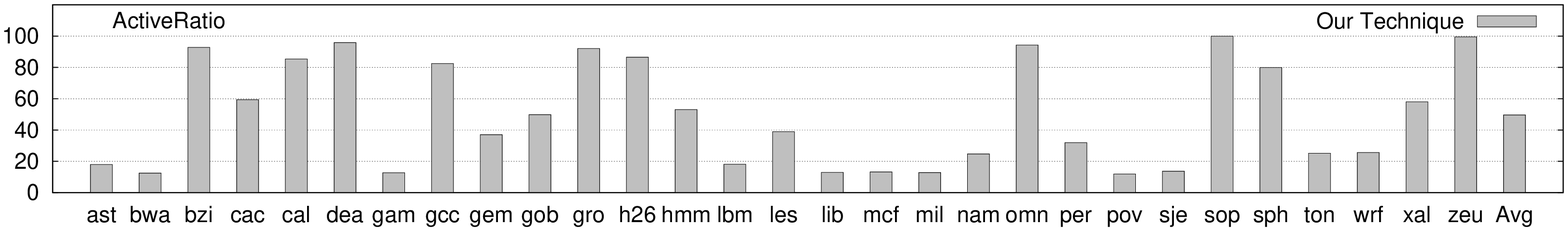}
\includegraphics [scale=0.45] {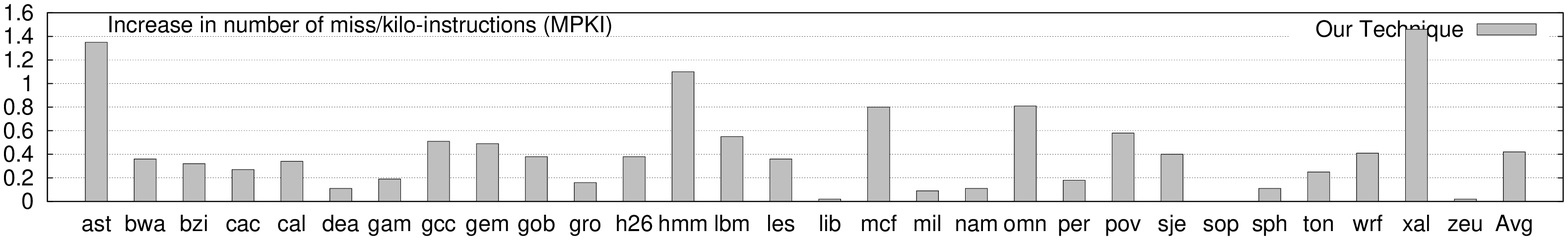}
\caption{Results for different schemes at 30 $\mu$s refresh period. Note that \% improvement in performance refers to \% reduction in simulation time, and hence, a higher value is better. Similarly, for decrease in RPKI and \% energy saved, a higher value is better.}
\label{fig:results30}
\end{figure*}

\section{Results and Analysis} \label{sec:results}

\subsection{Results with 40 $\mu$s Refresh Period}
Figure \ref{fig:results40} shows the results with a refresh period of 40 $\mu$s. On average, RPV saves 13.5\% energy, while our technique saves 22.8\% energy. Also, RPV improves performance by 5.9\%, while our technique improves performance by 9.7\%. Clearly, our technique outperforms RPV by a large margin, in terms of both performance and energy efficiency.  This is because of several reasons.   While RPV aims to save only refresh energy, our technique saves both leakage and refresh energy. While RPV keeps the entire cache ON, using our technique, the average ActiveRatio is 49.0\% and thus, our technique turns-off more than half the cache.  This also reflects in large saving in leakage energy. The average reduction in RPKI for RPV and our technique are 66.4 and 252.3, respectively, thus our technique reduces RPKI more effectively than RPV. Thus, even in terms of refresh energy, our technique provides larger savings than RPV. 

The average increase in MPKI by using our technique is 0.42, which is quite small. Thus, despite periodically reconfiguring the cache, our technique does not cause significant increase in off-chip traffic. 

As for SRAM cache, for all the applications, its energy efficiency is inferior to that of eDRAM baseline. On average, SRAM cache consumes 22.2\% more energy than the baseline eDRAM cache. This is because the leakage energy consumption of SRAM is significantly higher than that of the eDRAM. On taking SRAM cache as the baseline, the energy saving provided by our technique will be even \textit{larger}. This shows that effective management of leakage and refresh energy provided by our technique makes eDRAM cache a viable, energy-efficient alternative for large LLCs in modern processors.   SRAM cache provides better performance than the eDRAM baseline since it does not require refresh operations, which leads to saving of time. On average, SRAM cache has 19.5\% better performance than the baseline eDRAM cache.

\subsection{Results with 30$\mu$s Refresh Period}
Since retention period is expected to reduce with technological scaling \cite{changtechnology}, we test our technique with a refresh period of 30 $\mu$s.  Note that since SRAM does not require refresh operations, its results are unaffected by the refresh period. Figure \ref{fig:results30} shows the results with a refresh period of 30 $\mu$. 

 On average, RPV provides 16.0\% energy saving, while our technique provides     27.9\% energy saving. Thus, our technique provides large advantage over RPV. Compared to the case of 40$\mu$s refresh period, the energy savings of both the techniques is increased. This is because of the fact that for smaller refresh period, a larger fraction of energy is spent in the form of refresh energy and hence, the benefits provided by any refresh energy saving techniques are also increased. Similar explanation also applies for other metrics, viz. performance improvement and decrease in RPKI. 

On average, RPV provides 8.6\% and our technique provides 14.0\% better performance over the baseline eDRAM cache (see Figure \ref{fig:results30}). Further, RPV reduces RPKI by 103 and our technique reduces RPKI by 371, thus offering more than 3 times better advantage than the RPV technique. The average increase in MPKI by using our technique is 0.42, which is small.

As for SRAM cache, compared to eDRAM cache, it provides 2.3\% loss in energy. Comparing this with the case of 40$\mu$s refresh period, we observe that the energy advantage of eDRAM cache is reduced when the refresh period is smaller. Notice that even for a mere 10$\mu$s reduction in refresh period (from 40$\mu$s to 30$\mu$s) of eDRAM, its energy efficiency margin over SRAM cache is reduced from 22.2\% to 2.3\%. Thus, the energy advantage of a low-leakage device like eDRAM over conventional SRAM (and consequently, its adoption) will crucially depend on the refresh period.  Further, SRAM cache provides 26.2\% better performance than the eDRAM cache and with reduction in eDRAM refresh period, the performance advantage of SRAM is increased for the reasons mentioned above. 


The results presented in this section show that our technique is very effective in saving cache leakage and refresh energy. Saving in cache energy provided by our technique is also likely to reduce the temperature of the chip and cooling requirement \cite{patel2003smart} and hence, lead to further energy savings.


 \section{Conclusion}\label{sec:conclusion}
In this paper, we have presented a technique for saving leakage and refresh energy in eDRAM caches. The experimental results have shown that our technique provides significant energy savings compared to baseline eDRAM cache and also outperforms a state-of-the-art eDRAM energy saving technique. Our technique does not require offline profiling and hence, is suitable for real-world systems which execute trillions of instructions. Our future work will focus on evaluating our technique for wide range of system parameters and for multicore systems.



\ifCLASSOPTIONcaptionsoff
  \newpage
\fi



\bibliographystyle{IEEEtran}
\bibliography{PhDReferences}

\begin{thebibliography}{10}
\providecommand{\url}[1]{#1}
\csname url@samestyle\endcsname
\providecommand{\newblock}{\relax}
\providecommand{\bibinfo}[2]{#2}
\providecommand{\BIBentrySTDinterwordspacing}{\spaceskip=0pt\relax}
\providecommand{\BIBentryALTinterwordstretchfactor}{4}
\providecommand{\BIBentryALTinterwordspacing}{\spaceskip=\fontdimen2\font plus
\BIBentryALTinterwordstretchfactor\fontdimen3\font minus
  \fontdimen4\font\relax}
\providecommand{\BIBforeignlanguage}[2]{{%
\expandafter\ifx\csname l@#1\endcsname\relax
\typeout{** WARNING: IEEEtran.bst: No hyphenation pattern has been}%
\typeout{** loaded for the language `#1'. Using the pattern for}%
\typeout{** the default language instead.}%
\else
\language=\csname l@#1\endcsname
\fi
#2}}
\providecommand{\BIBdecl}{\relax}
\BIBdecl

\bibitem{4554309}
A.~Pande, A.~Mittal, A.~Verma, and P.~Kumar, ``Meeting real-time requirements
  for a low bitrate multimedia encoding framework,'' in \emph{IEEE
  International Conference on Electro/Information Technology}, 2008, pp.
  258--262.

\bibitem{Mit_DRAMsurvey}
S.~Mittal, ``{A Survey of Architectural Techniques For DRAM Power
  Management},'' \emph{International Journal of High Performance Systems
  Architecture}, vol.~4, no.~2, pp. 110--119, 2012.

\bibitem{pande2013demonstration}
A.~Pande \emph{et~al.}, ``Accurate energy expenditure estimation using
  smartphone sensors,'' \emph{ACM Wireless Health Conference}, 2013.

\bibitem{rodriguez2006energy}
S.~Rodriguez and B.~Jacob, ``Energy/power breakdown of pipelined nanometer
  caches (90nm/65nm/45nm/32nm),'' in \emph{International symposium on Low power
  electronics and design (ISLPED)}, 2006, pp. 25--30.

\bibitem{monchiero2006design}
M.~Monchiero, R.~Canal, and A.~Gonz{\'a}lez, ``Design space exploration for
  multicore architectures: a power/performance/thermal view,'' in
  \emph{ICS}.\hskip 1em plus 0.5em minus 0.4em\relax ACM, 2006, pp. 177--186.

\bibitem{li2009mcpat}
S.~Li, J.~Ahn, R.~Strong, J.~Brockman, D.~Tullsen, and N.~Jouppi, ``{McPAT: an
  integrated power, area, and timing modeling framework for multicore and
  manycore architectures},'' in \emph{42nd IEEE/ACM International Symposium on
  Microarchitecture (MICRO)}, 2009, pp. 469--480.

\bibitem{kalla2010power7}
R.~Kalla, B.~Sinharoy, W.~J. Starke, and M.~Floyd, ``{Power7: IBM's
  next-generation server processor},'' \emph{Micro, IEEE}, vol.~30, no.~2, pp.
  7--15, 2010.

\bibitem{iyer2005embedded}
S.~Iyer, J.~Barth~Jr, P.~Parries, J.~Norum, J.~Rice, L.~Logan, and D.~Hoyniak,
  ``{Embedded DRAM: Technology platform for the Blue Gene/L chip},'' \emph{IBM
  Journal of Research and Development}, vol.~49, no. 2.3, pp. 333--350, 2005.

\bibitem{barth2008500}
J.~Barth, W.~R. Reohr, P.~Parries, G.~Fredeman, J.~Golz, S.~E. Schuster, R.~E.
  Matick, H.~Hunter, C.~C. Tanner, J.~Harig \emph{et~al.}, ``{A 500 MHz random
  cycle, 1.5 ns latency, SOI embedded DRAM macro featuring a three-transistor
  micro sense amplifier},'' \emph{IEEE Journal of Solid-State Circuits},
  vol.~43, no.~1, pp. 86--95, 2008.

\bibitem{wilkerson2010reducing}
C.~Wilkerson, A.~R. Alameldeen, Z.~Chishti, W.~Wu, D.~Somasekhar, and S.-l. Lu,
  ``Reducing cache power with low-cost, multi-bit error-correcting codes,''
  \emph{ACM SIGARCH Computer Architecture News}, vol.~38, no.~3, pp. 83--93,
  2010.

\bibitem{changtechnology}
M.-T. Chang, P.~Rosenfeld, S.-L. Lu, and B.~Jacob, ``{Technology Comparison for
  Large Last-Level Caches (L$^3$Cs): Low-Leakage SRAM, Low Write-Energy
  STT-RAM, and Refresh-Optimized eDRAM},'' \emph{International Symposium on
  High-Performance Computer Architecture (HPCA)}, 2013.

\bibitem{agrawalrefrint}
A.~Agrawal, P.~Jain, A.~Ansari, and J.~Torrellas, ``Refrint: Intelligent
  refresh to minimize power in on-chip multiprocessor cache hierarchies,''
  \emph{International Symposium on High-Performance Computer Architecture
  (HPCA)}, 2013.

\bibitem{kumar2011distributed}
P.~Kumar \emph{et~al.}, ``Distributed video coding and content analysis for
  resource constraint multimedia applications,'' \emph{Ubiquitous Multimedia
  and Mobile Agents: Models and Implementations}, p. 251, 2011.

\bibitem{mittal2010content}
A.~Mittal \emph{et~al.}, ``Content-based network resource allocation for real
  time remote laboratory applications,'' \emph{Signal, Image and Video
  Processing}, vol.~4, no.~2, pp. 263--272, 2010.

\bibitem{pande2010algorithms}
A.~Pande, ``Algorithms and architectures for secure embedded multimedia
  systems,'' Ph.D. dissertation, Iowa State University, 2010.

\bibitem{MitMit2011_QA}
S.~Mittal and A.~Mittal, ``{Versatile question answering systems: seeing in
  synthesis},'' \emph{International Journal of Intelligent Information and
  Database Systems}, vol.~5, pp. 119--142, 2011.

\bibitem{pande2007network}
A.~Pande, A.~Verma, A.~Mittal, and A.~Agrawal, ``Network aware efficient
  resource allocation for mobile-learning video systems,'' in \emph{Proc. Intl.
  Conf. Mobile Learning}, 2007, pp. 189--196.

\bibitem{ghosh2007smart}
M.~Ghosh and H.-H.~S. Lee, ``{Smart refresh: An enhanced memory controller
  design for reducing energy in conventional and 3D Die-Stacked DRAMs},'' in
  \emph{40th Annual IEEE/ACM International Symposium on Microarchitecture},
  2007, pp. 134--145.

\bibitem{reohr2006memories}
W.~R. Reohr, ``Memories: Exploiting them and developing them,'' in \emph{IEEE
  International SOC Conference}, 2006, pp. 303--310.

\bibitem{wu2009hybrid}
X.~Wu \emph{et~al.}, ``Hybrid cache architecture with disparate memory
  technologies,'' in \emph{ACM SIGARCH Computer Architecture News}, vol.~37,
  no.~3, 2009, pp. 34--45.

\bibitem{HanHri02_TVLSI}
H.~Hanson, M.~Hrishikesh, V.~Agarwal, S.~Keckler, and D.~Burger, ``Static
  energy reduction techniques for microprocessor caches,'' \emph{IEEE
  Transactions on VLSI Systems}, vol.~11, no.~3, pp. 303 --313, 2003.

\bibitem{YanPow01_IcacheResize}
S.-H. Yang, B.~Falsafi, M.~D. Powell, K.~Roy, and T.~N. Vijaykumar, ``An
  integrated circuit/architecture approach to reducing leakage in
  deep-submicron high-performance {I}-caches,'' in \emph{7th International
  Symposium on High-Performance Computer Architecture (HPCA)}, 2001, pp. 147--.

\bibitem{mittal2013PhDThesis}
S.~Mittal, ``Dynamic cache reconfiguration based techniques for improving cache
  energy efficiency,'' Ph.D. dissertation, Iowa State University, 2013.

\bibitem{KesHil92_PageColoring}
R.~Kessler and M.~Hill, ``Page placement algorithms for large real-indexed
  caches,'' \emph{ACM Transactions on Computer Systems (TOCS)}, vol.~10, no.~4,
  pp. 338--359, 1992.

\bibitem{mittal2013cache}
S.~Mittal, \emph{Cache Energy Optimization Techniques for Modern
  Processors}.\hskip 1em plus 0.5em minus 0.4em\relax Scholars' Press, 2013.

\bibitem{puzaksampling}
T.~Puzak, ``Cache memory design,'' Ph.D. dissertation, University of
  Massachusetts, 1985.

\bibitem{EyeEec06_CPIOutOrder}
S.~Eyerman, L.~Eeckhout, T.~Karkhanis, and J.~E. Smith, ``{A performance
  counter architecture for computing accurate CPI components},'' in \emph{12th
  International Conference on Architectural Support for Programming Languages
  and Operating Systems (ASPLOS)}, 2006, pp. 175--184.

\bibitem{CarHei2011_Sniper}
T.~E. Carlson, W.~Heirman, and L.~Eeckhout, ``Sniper: Exploring the level of
  abstraction for scalable and accurate parallel multi-core simulations,'' in
  \emph{International Conference for High Performance Computing, Networking,
  Storage and Analysis (SC)}, Nov. 2011.

\bibitem{chang200765nm}
M.-T. Chang, P.-T. Huang, and W.~Hwang, ``{A 65nm low power 2T1D embedded DRAM
  with leakage current reduction},'' in \emph{IEEE International SOC
  Conference}, 2007, pp. 207--210.

\bibitem{chun2010logic}
K.~C. Chun, P.~Jain, and C.~H. Kim, ``{Logic-compatible embedded DRAM design
  for memory intensive low power systems},'' in \emph{IEEE International
  Symposium on Circuits and Systems (ISCAS)}, 2010, pp. 277--280.

\bibitem{cacti_53}
\mbox{CACTI} 5.3, \url{http://quid.hpl.hp.com:9081/cacti/}, 2013.

\bibitem{patel2003smart}
C.~D. Patel, C.~E. Bash, R.~Sharma, M.~Beitelmal, and R.~Friedrich, ``Smart
  cooling of data centers,'' \emph{Pacific RIM/ASME International Electronics
  Packaging Technical Conference and Exhibition (IPACK03)}, 2003.

\end{thebibliography}

\end{document}